\documentclass[12pt]{iopart}

\expandafter\let\csname equation*\endcsname\relax
\expandafter\let\csname endequation*\endcsname\relax
\usepackage{amsmath, iopams}
\usepackage{slashed}

\usepackage[margin=0.5cm]{caption}
\usepackage{tikz}\usetikzlibrary{positioning}
\usepackage{graphics}

\usepackage{cancel}
\usepackage{amsfonts,amssymb}
\def\rvarphi{\raise 2pt\hbox{$\varphi$}}
\usepackage{graphicx}
\usepackage{float}
\usepackage{subfig}
\usepackage{overpic}
\usepackage{hyperref}
\usepackage{doi}

\begin{document}
\title[Angular velocities of rotating black holes]{Angular velocity of rotating black holes - a new way to construct initial data for binary black holes}

\author{Shuanglin Huang$^{1,2}$\footnote{Joint first author.}, Xuefeng Feng$^{3,4}$\footnote{Joint first author.} 
Yun-Kau Lau$^{5}$\footnote{Corresponding author, email: lau@amss.ac.cn.}}
\address{$^1$Beijing Institute of Mathematical Sciences and Applications, Yanqihu, Huairou District, Beijing, 101408, China}
\address{$^2$Yau Mathematical Sciences Center, Tsinghua University, Beijing, 100084, China}
\address{$^3$Fudan Center for Mathematics and Interdisciplinary Study, Fudan University, Shanghai, 200433, China}
\address{$^4$Shanghai Institute for Mathematics and Interdisciplinary Sciences (SIMIS), Shanghai, 200433, China} 
\address{$^5$Institute of Applied Mathematics, Morningside Center of Mathematics, Academy of Mathematics and System Science, Chinese Academy of Sciences, 55, Zhongguancun Donglu, Beijing, 100190, China}

\begin{abstract}
Motivated by a geometric understanding of the angular velocity of a Kerr black hole in terms of a quasi-conformal map that describes a 2d Beltrami fluid flow, a new way to construct initial data sets for binary rotating black holes by prescribing the angular velocities of the two black holes at their horizons is discussed. 
A set of elliptic equations with prescribed Dirichlet boundary conditions at the horizons and at spatial infinity is established for constructing the initial data. To explore the dynamics encoded in these initial data, we consider the conformally flat three-metric case and numerically evolve it using the BSSN code for two co-rotating and counter-rotating black holes with angular velocities prescribed at the horizons. When the angular velocities are non-uniform and deviate from a constant value at the horizons, 
new gravitational waveforms are generated which display certain oscillatory pattern reminiscent of that of quasi-normal ringing in the inspiral phase before merger takes place.

\end{abstract}

\maketitle

\section{Introduction}

The aim of the present work is to propose a new method to construct initial data sets for the vacuum Einstein field equations by prescribing  the dynamics of rotation at the black hole boundary, motivated primarily by our recent effort to understand the geometry of angular velocity of the Kerr metric\cite{feng2025geometric}. For a constant time slicing of the Kerr metric, the momentum constraint can be written as a Beltrami equation\cite{kuhnau2002handbook}, which appears in the context of two-dimensional fluid flow.The Beltrami equation is a fundamental tool in the study of quasi-conformal mappings and complex analysis. Furthermore, it offers a powerful framework for describing specific structures in two-dimensional fluid dynamics. The classical Beltrami equation governs a complex-valued function $f$ through the relation
\[
\frac{\partial f}{\partial \bar{z}} = \mu(z) \frac{\partial f}{\partial z}, 
\qquad |\mu(z)| < 1,
\]
where the function $\mu$, known as the Beltrami coefficient, encapsulates the degree of non-conformality. A potential function can then be constructed for the momentum constraint and identified as the angular velocity of the Kerr metric. Since the construction is local, this fluid dynamical picture may be generalized to multiple black holes, in particular to binary black holes. Within the context of binary black holes, boundary conditions are prescribed at the apparent horizons in accordance with the rotational dynamics of these black holes. This enables us to establish a link between the spacetime dynamics in the vicinity of the horizons and the gravitational waveform extracted near null infinity in the time evolution of the initial data sets. Hopefully it offers an avenue to explore the dynamics of gravity near the horizon in relation to the gravitational wave form and gravitational wave will become a probe of the near horizon dynamics of binary black holes. For black holes with non-uniform angular velocities, as we shall see, the inspiral phase of the waveforms 
exhibits certain new features not seen before. Details of the construction are presented in the ensuing sections.

The paper is structured as follows. In Section II, we briefly describe the geometric description of the angular velocity of a  Kerr black hole in terms of a quasi-conformal map. 
In Section III, within the conformally flat context, we build up a framework for the initial data construction by prescribing the angular velocity of a black hole at the horizon. In Section IV, by means of numerical simulation, we try to understand the gravitational waveform generated by the newly constructed initial data, to be followed by some concluding remarks in section V. 

\section{Angular velocity of a rotating black hole and the Kerr metric}

In this section, we provide a brief geometric description of the angular velocity in the Kerr metric in terms of the momentum constraint on a constant time slice. This will pave the way for constructing initial data in terms of the angular velocities of binary black holes in what follows. 
 
Consider the Kerr metric in the Boyer-Lindquist coordinates $(t,r,\theta,\phi)$, given by 
\begin{eqnarray}\label{eq:kerrmetric}
ds^2&=&-\left (1-\frac{2Mr}{\Sigma}\right)\ dt^2-\frac{4aMr {\rm sin}^2\theta}{\Sigma}\ dtd\phi \nonumber\\
 & &+ \frac{\Sigma}{\Delta}\ dr^2  + \Sigma\ d\theta^2 + \frac{A}{\Sigma}\sin^2\theta d\phi^2\,,
\end{eqnarray}
where $A=(r^2+a^2)^2-\Delta a^2\sin^2\theta$, $\Sigma=r^2+a^2\cos^2\theta$ and $\Delta=r^2-2Mr+a^2$.

Let $K_{ab}$ denote the extrinsic curvature of the constant time slice, where the maximal slicing condition $K=0$ is satisfied. 
Consider the vector field
\begin{eqnarray}\label{eq:vector}
    V_{a}=K_{ab}\eta^{b}\,,
\end{eqnarray}
where $\eta^{a}=X^{-1/2}\phi^{a}$ is the normalized axisymmetric Killing vector and $X$ is the norm of the axisymmetric Killing vector field $\phi^a$.
Further define $v_a=XV_a$. It may be inferred from the momentum constraint equation and the axisymmetric structure of $V_a$ that, in terms of the flat connection $\partial_a$ on the two-plane
orthogonal to $\phi_a$, 
the following Hodge-like system is satisfied. 
\begin{eqnarray}\label{continuity}
    \partial^{a}v_{a}=0\,,
\end{eqnarray}
\begin{eqnarray}
     \partial_{[a} \lambda X^{-\frac{3}{2} }v_{b]}=0\,,\label{continuity2}
\end{eqnarray}
where $\lambda$ is the stationary Killing field for the constant time slicing. Equations (\ref{continuity}) and (\ref{continuity2}) resemble the Beltrami flow for a compressible fluid in the two-plane orthogonal to $\phi_a$. A quasi-conformal equation
can be further developed from (\ref{continuity}) and (\ref{continuity2}) as 
\begin{eqnarray}
    (w\,\Omega_{x})_{x}+(w\,\Omega_{y})_{y}=0\,,\label{Omega-eq}\\
     (w\,\Psi_{x})_{x}+(w\,\Psi_{y})_{y}=0\,.\nonumber
	\end{eqnarray}
$\Omega$ may be identified as the angular velocity of a rotating black hole, originally defined in terms of the stationary and axisymmetric Killing field. $\Psi$
may be regarded as a stream function conjugate to $\Omega$.

\begin{figure*}[htbp]
	\centering	\includegraphics[width=0.6\textwidth]{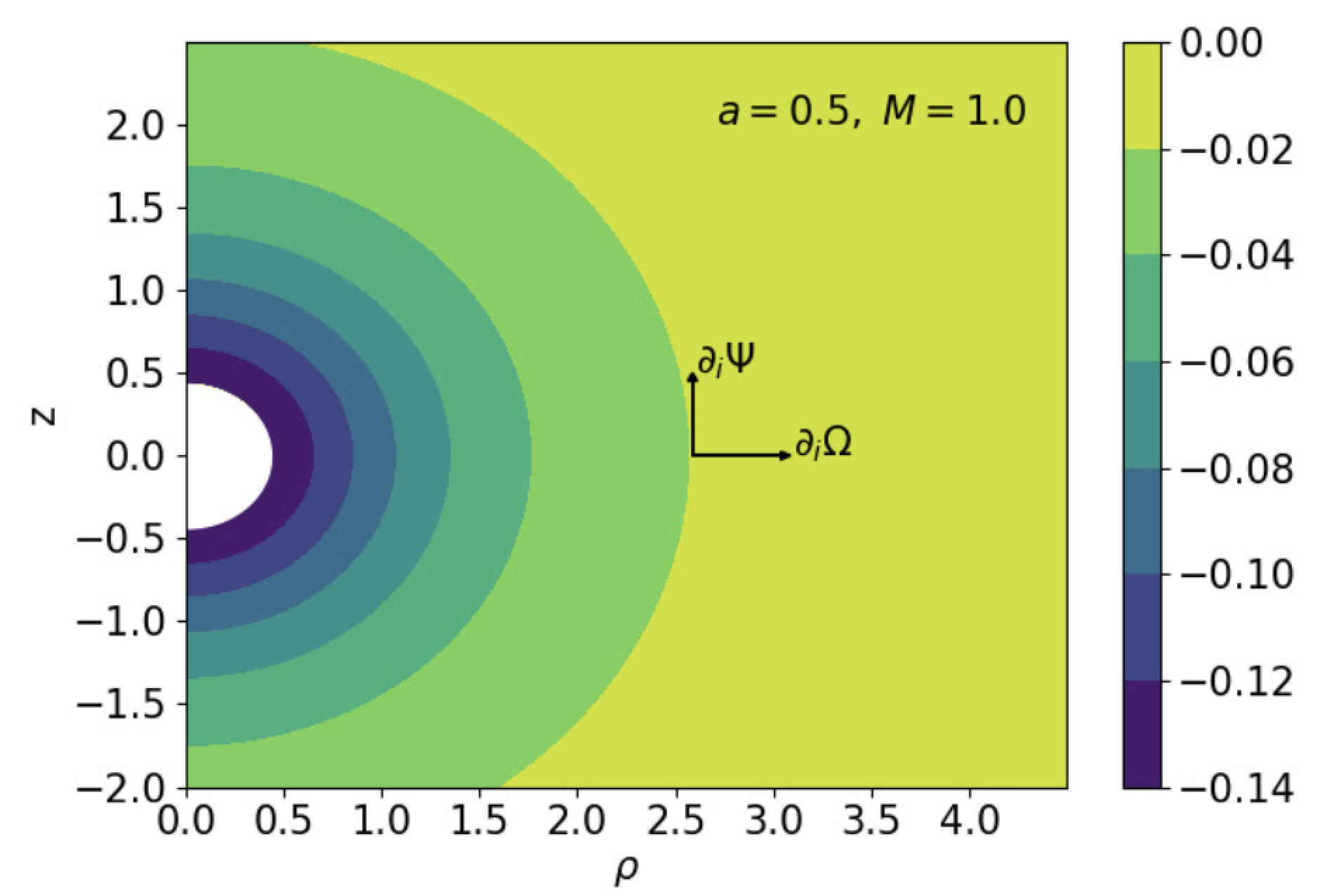}
	\caption{A Beltrami flow defined near the horizon by $\Omega$ and its conjugate stream function $\Psi$. }\label{fig:fluid}
\end{figure*}

\section{Angular velocities of binary black holes as initial data }
When expressed in Weyl coordinates, the initial data set for the Kerr metric described in the preceding section suggests a very natural generalization to the case of Brill waves for binary rotating black holes, where the angular velocities of the black holes are treated as part of the initial data. As a first step toward understanding the construction of these new initial data sets, we shall consider a simple setting: a conformally flat initial data slice. The Bowen-York data are then employed as a benchmark for comparison with our construction.  

Consider the conformally flat three-metric
\begin{eqnarray}\label{eq:cylin2}
   \gamma_{ij}\,dx^idx^j=\psi^4(d\rho^2+dz^2+\rho^2d\phi^2)\,,
\end{eqnarray}
where the flat metric is given in terms of standard cylindrical coordinates. Let $K_{ij}$ denote the extrinsic curvature of the slice, and impose the maximal-slicing condition $K=0$ on $K_{ij}$. Define tensor $\bar{K}_{ij}$ as:
\begin{eqnarray}
    K_{ij}&=\psi^{-2}\bar K_{ij}\,.
\end{eqnarray}
The constraint equations are then given by 
\begin{eqnarray}
  \bar{D}^2\psi+\frac{1}{8}\bar{K}_{ij}\bar{K}^{ij}\psi^{-7}&=0\,,\label{eq:ham1}\\
  \bar{D}_j\bar{K}^{ij} = 0\,,\label{eq:mom1}
\end{eqnarray}
where $\bar D$ is defined as the covariant derivative operator with respect to the flat metric.

Let $\bar K^\Omega_{ij}$ be $\bar K_{ij}$ subject to the momentarily stationary slicing condition (or equivalently, the $(t, \phi)$-symmetric condition), such that 
\begin{eqnarray}
\bar{K}^{\tiny\Omega}_{ij}h^{im}h^{jn}=0\,,\quad \bar{K}^{\Omega}_{ij}\phi^{i}\phi^{j}=0\,. \label{tphisym}
\end{eqnarray}
Here $h_{ij}$ is the two-metric induced by restricting $\gamma_{ij}$ to the $(\rho, z)$ plane, and $\phi^i$ denotes the azimuthal Killing vector $\left(\partial/\partial \phi\right)^i$.
$\bar{K}^{\Omega}_{\rho\phi}$ and $\bar{K}^{\Omega}_{z\phi}$ are the only two independent, non-zero components of $\bar{K}^{\Omega}_{ij}$. 
Further define the vector field $\bar{V}_{a}$
\begin{eqnarray}
\bar{V}_{i}=\bar{K}^{\Omega}_{ij}\bar{\eta}^{j}\,,
\end{eqnarray}
where 
\begin{eqnarray}    \bar{\eta}^{i}=\rho^{-1}\phi^{i}\,.\nonumber
\end{eqnarray}
As in the case of the Kerr metric, 
the vector field $\bar{V}_{i}$ may be expressed by the angular velocity potential $\Omega$ as,
\begin{eqnarray}
    \bar{V}_{i}=-\frac{1}{2}\rho\partial_{i}\Omega\,.
\end{eqnarray}
where we have chosen the lapse function to be $\alpha=\psi^6$. $\bar{K}^{\Omega}_{ij}$ can be constructed from $\Omega$ as:
\begin{eqnarray}\label{eq:Omega}
  \bar{K}^{\Omega}_{ij} = \bar{\eta}_{(i} \bar{V}_{j)} = -\frac{1}{2}\rho \bar{\eta}_{(i} \partial_{j)}\Omega\,.
\end{eqnarray}
Given Eq.\eqref{eq:Omega}, the Hamiltonian and momentum constraint equations can be expressed as elliptic equations for the conformal factor $\psi$ and the potential $\Omega$:
\begin{eqnarray}
    \bar{D}^2\psi+\frac{1}{8}\bar{K}_{ij}\bar{K}^{ij}\psi^{-7}=0\,,\label{eq:ham2}\\
    \partial^{i}(\rho^{3}\partial_{i}\Omega)=0\,.\label{eq:mom2}
\end{eqnarray}

As it stands, axisymmetry of \eqref{eq:Omega} means that the initial data constructed can only describe head-on collision of two black holes. To incorporate inspiral in the data, linearity of the momentum constraint \eqref{eq:mom1} enables us to supplement \eqref{eq:mom2} by adopting the linear momentum part of the Bowen-York ansatz into the initial data set. Specifically, we set:
\begin{eqnarray}
   \bar{K}_{ij} = \bar{K}_{ij}^{P} + \bar{K}_{ij}^{\Omega}\,.
\end{eqnarray}
in which
\begin{eqnarray}
  \bar{K}_{ij}^{P}&=\frac{3}{2r_{A}^2}\big[P_{Ai}n_{Aj}+P_{Aj}n_{Ai}-(\delta_{ij}-n_{Ai}n_{Aj})P^{k}_{A}n_{Ak}\big]\,,\\ \label{eq:KP}
 \bar{K}^{\Omega}_{ij} &= -\frac{1}{2}\rho \bar{\eta}_{(i} \partial_{j)}\Omega\,.
\end{eqnarray}
Here $P^{i}_{A}$ are constant vectors, and $r_A=\Vert x^i-c^i_A\Vert$ is the coordinate distance to the center of black hole located at $x^i = c^i_A$, and $n^i_A = (x^i-c^i_A)/r_A$. $\bar K_P^{ij}$ is the linear momentum part of the Bowen-York solution \cite{bowen1980time}, in which $P_A^i$ can be loosely regarded as the linear momentum of each black hole.

The spin part of the Bowen-York solution, which is not adopted in our method, is
\begin{eqnarray}
    \bar{K}_{ij}^{S}&=\frac{3}{r_{A}^3}(n_{Ai}\epsilon_{jkl}+n_{Aj}\epsilon_{ikl})S_{A}^k n_{A}^l\,,
\end{eqnarray}
where constant vector $S_A^i$ can be associated with the spin of each black hole in the limit of infinite separation. $\bar{K}_{ij}^{S}$ also satisfies condition \eqref{tphisym}, and therefore it corresponds to a special choice of $\Omega$ in our method. 
Without loss of generality, we assume one black hole Bowen-York data with angular momentum directed along the $z$ axis, i.e.: 
\begin{eqnarray}
S^i &= (0,J,0)\,,\\
n^i &= \frac{1}{(\rho^2+z^2)^{\frac{1}{2}}}(\rho,z,0)\,.
\end{eqnarray}
then the non-zero terms of $\bar{K}^S_{ij}$ are:
 \begin{eqnarray}
\bar{K}^S_{\rho\phi}=\bar{K}^S_{\phi\rho}=\frac{3J\rho^3}{(\rho^2+z^2)^{\frac{5}{2}}}\,,\\ \bar{K}^S_{z\phi}=\bar{K}^S_{\phi z}=\frac{3J\rho^2z}{(\rho^2+z^2)^{\frac{5}{2}}}\,, 
\end{eqnarray}
It is straightforward to verify that this corresponds to  
\begin{eqnarray}
\Omega=\frac{J}{(\rho^2+z^2)^{\frac{3}{2}}}\,.\label{exp:omega_J}
\end{eqnarray}
in $\bar{K}^{\Omega}_{ij}$. This form of $\Omega$ corresponds to a particular spin distribution: it describes an axisymmetric rotation with total angular momentum $J$ aligned along the $z$-axis, in which the angular velocity is uniform on spheres of constant $r=\sqrt{\rho^2+z^2}$.

\subsection{Boundary Conditions}

The initial data are required to be asymptotically flat, and the outer boundary $\mathcal{B}_{\infty}$ is placed near infinity. 
The outer boundary condition for $\psi$ is
\begin{eqnarray} 
    \psi=1\quad {\rm on }\ \mathcal{B}_{\infty}\,.
\end{eqnarray}
In the case of the Kerr metric, the angular velocity $\Omega=\mathcal{O}(r^{-3})$ when $r\rightarrow\infty$. It is reasonable to adopt a similar asymptotic behavior as the boundary condition for $\Omega$ at infinity in the binary case.

To specify boundary conditions for black hole initial data, we require the inner boundary to consist of marginally trapped surfaces. This leads to the following boundary condition:
\begin{eqnarray}
    \frac{\partial\psi}{\partial r} + \frac{\psi}{2r} + \frac{\bar{K}_{rr}}{4\psi^3} = 0\,.\label{boundary_eq}
\end{eqnarray}
Each trapped surface is chosen to be a coordinate two-sphere located at a constant radius $r$, where $r$ is the radial coordinate centered on each black hole.
It should be noted that this condition does not guarantee that the inner boundary coincides with the apparent horizon, which is defined as the outermost marginally trapped surface. However, in practice, once the initial data have been constructed, the apparent horizon can always be located as the outermost such surface. 

In our numerical setup, we enforce the condition \eqref{boundary_eq} as follows. We first prescribe
$\Omega$ at the inner boundary according to the desired configuration, then solve for $\Omega$ and combine it with the
specified $\bar{K}^P_{ij}$ to construct $\bar{K}_{ij}$. Condition \eqref{boundary_eq} is subsequently used as a boundary condition for solving the conformal factor $\psi$.

\section{Waveform simulations}\label{sec:sim} 

In order to understand the gravitational wave content of the newly constructed data, we present in this section the gravitational waveforms generated by the evolution of black hole initial data sets constructed using our method. To facilitate comparison with the Bowen-York initial data, we adjust the free parameters such that the Christoudoulou mass and spin of black holes are identical to those constructed using Bowen-York method. 

\subsection{Numerical settings}
To construct the initial data, we developed an elliptic solver based on LORENE\cite{Lorene:web}, a C++ library provides tools for solving PDEs using multi-domain spectral methods. We adopt the treatment in \cite{gourgoulhon2001quasiequilibrium, grandclement2002binary}, space is decomposed into various spherical-like shells, and the outermost shell is compactified using the variable $u = 1/r$, which enables one to impose exact boundary conditions at infinity. For binary black holes, two sets of such domains are used. Each set is centered on one black hole, and the fields are decomposed into contributions defined on the two sets of domains.

The Einstein Toolkit \cite{loffler2012einstein, EinsteinToolkit:2025_05} is employed to simulate the evolution of black hole collisions and inspirals. For comparison, we also evolve standard Bowen-York initial data constructed using the \texttt{TwoPunctures}\cite{Ansorg:2004ds} thorn. Both sets of initial data are evolved using the \texttt{McLachlan/ML\_BSSN}\cite{Brown:2008sb, McLachlan:web} thorn, which implements the BSSN formulation\cite{shibata1995evolution,baumgarte1998numerical} of Einstein’s equations, together with the \texttt{Carpet}\cite{	
Schnetter:2003rb} driver for adaptive mesh refinement. The apparent horizons of the black holes are located and tracked using the \texttt{AHFinderDirect}\cite{Thornburg:2003sf} thorn, and their quasi-local properties, such as mass and spin, are computed with the \texttt{QuasiLocalMeasures}\cite{Dreyer:2002mx} thorn.
The gravitational wave signal is extracted via the \texttt{WeylScal4} thorn.

\subsection{Single spinning black hole}
We first look at the single spinning black hole. Black hole spins are commonly characterized in terms of their dimensionless spin
\begin{align}
    \chi=\frac{S}{M_{\rm Chr}^2}\,.
\end{align}
Here $S$ is taken to be non-negative and represents a suitable
quasilocal spin, while $M_{\rm Chr}$ denotes the Christoudoulou mass \cite{christodoulou1970reversible} of the black hole, given by
\begin{align}
    M_{\rm Chr}^2=M_{\rm irr}^2+\frac{S^2}{4M_{\rm irr}^2}\,.
\end{align}
The irreducible mass $M_{\rm irr}$ is defined in terms of the area of the black hole's apparent horizon:
\begin{align}
    M_{\rm irr}=\sqrt{\frac{A}{16\pi}}.
\end{align}
Using a surface integral over the apparent horizon $H$, the Komar angular momentum can be calculated as follows\cite{brown1993quasilocal,ashtekar2001mechanics,ashtekar2003dynamical}:
\begin{align}
    S=\frac{1}{8\pi}\oint_{H}\phi^{i}s^{j}K_{ij}dA=\frac{1}{8\pi}\oint_{H}(\psi^4\rho^2)^{1/2}s^{i}V_{i}dA\,,
\end{align}
where $H$ denotes the black hole's apparent horizon, $s^{j}$ is the outward-pointing unit normal to $H$ within the $t=const$ hypersurface, and $\phi^{i}$ is an azimuthal vector field tangent to $H$.

\begin{figure}[h]
    \centering
    \includegraphics[height=6.0cm,width=8cm]{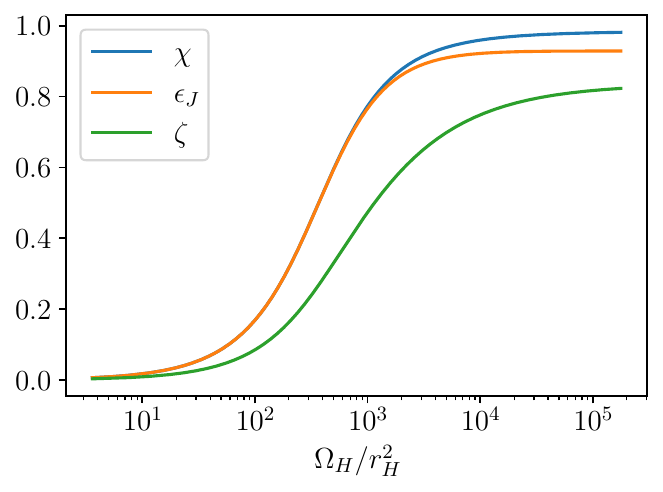} 
    \caption{The three spin parameters $\chi$, $\epsilon_{J}$, and $\zeta$ for a single rotating black hole are plotted, where a
value of zero denotes a non-spinning black hole and a value of one corresponds
to an extremal case. As the value of $\Omega_{H}/r_{H}^2$ increases, the parameters approach the following asymptotic values: $\chi_{\rm max}=0.984$, $\zeta_{\rm max}=0.823$, and $\epsilon_{J,\rm{max}}(S/M_{\rm ADM}^2)=0.928$.}
    \label{single_black_hole}
\end{figure}

Fig.~\ref{single_black_hole} shows the results for a single spinning black hole with inner boundary value of $\Omega$ prescribed as $\Omega|_{r=r_H}=\Omega_H \sin\theta$, where $\Omega_H$ is a constant parameter. In the plot, in addition to dimensionless spin $\chi$, we also introduce two other dimensionless spin measures, $\epsilon$ and $\zeta$, defined as
\begin{align}
    \epsilon=\frac{J_{ADM}}{E_{ADM}}\,,\,\,\, \zeta=\frac{S}{2M_{\rm irr}^2}\,.
\end{align}
$J_{ADM}$ and $E_{ADM}$ denote ADM angular momentum and ADM energy, respectively.  These three quantities $\chi$, $\epsilon_{J}$, and $\zeta$ are used to monitor whether a black hole becomes super-extremal, and it was observed that none of them reach or exceed unity as $\Omega_{H}/r_{H}^2$ increases. These findings are in good agreement with the results reported in the literature \cite{dain2002new,lovelace2008binary}.

\subsection{Constant $\Omega$ on the horizon and code calibration}
 For binary black holes, we first simulate the waveform with a constant value of $\Omega$ on the horizon. The waveform generated turns out to be nearly identical to that of the Bowen-York case as one would expect. This is consistent with the calculations in the previous section. This serves to calibrate our code so that, for the waveform generated by a more complicated choice of $\Omega$ which describes non-uniform angular velocity on the horizon in what follows, the deviation of waveform from that of Bowen-York is not due to numerical errors.

\subsection{Head-on collisions}

\begin{figure*}[ht!]
    \centering
    \includegraphics[height=5cm]{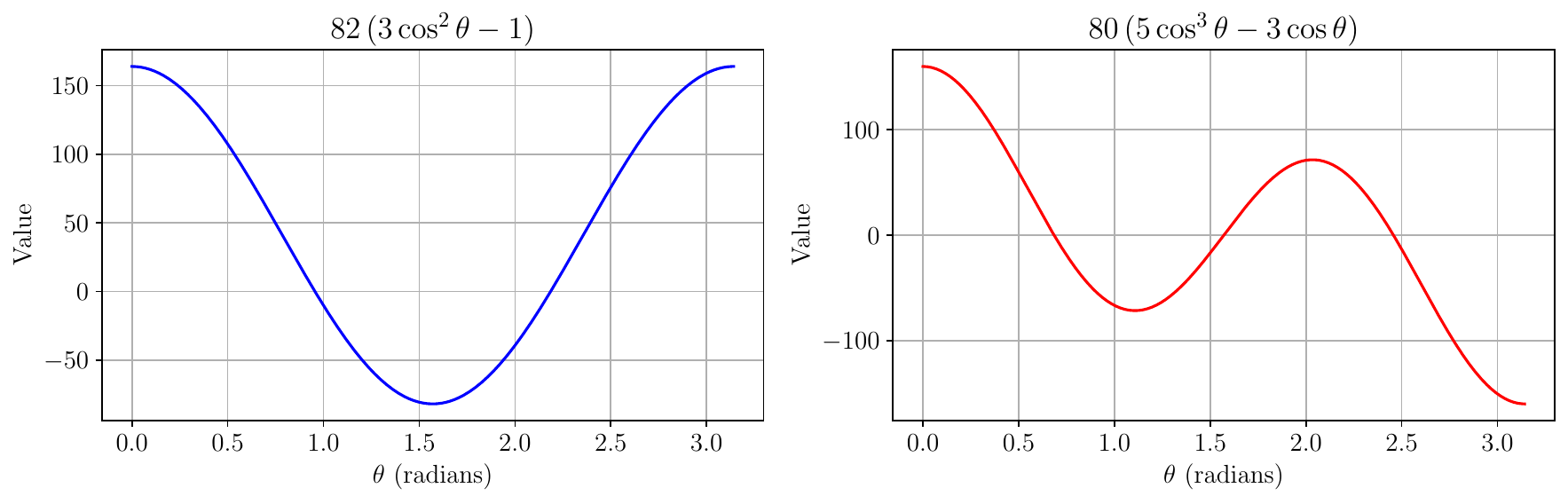} 
    \caption{Angular velocity function $\Omega(\theta)$ evaluated at the horizon. The profile in the left panel corresponds to the models in Fig.~\ref{fig_headon}, \ref{Inspiral_aligned_20}, and \ref{Inspiral_anti_20}, while the one in the right panel is employed in Fig.~\ref{Inspiral_30}.}\label{C20_C30}
\end{figure*}

\begin{figure*}[ht!]
    \centering
    \includegraphics[height=5.5cm]{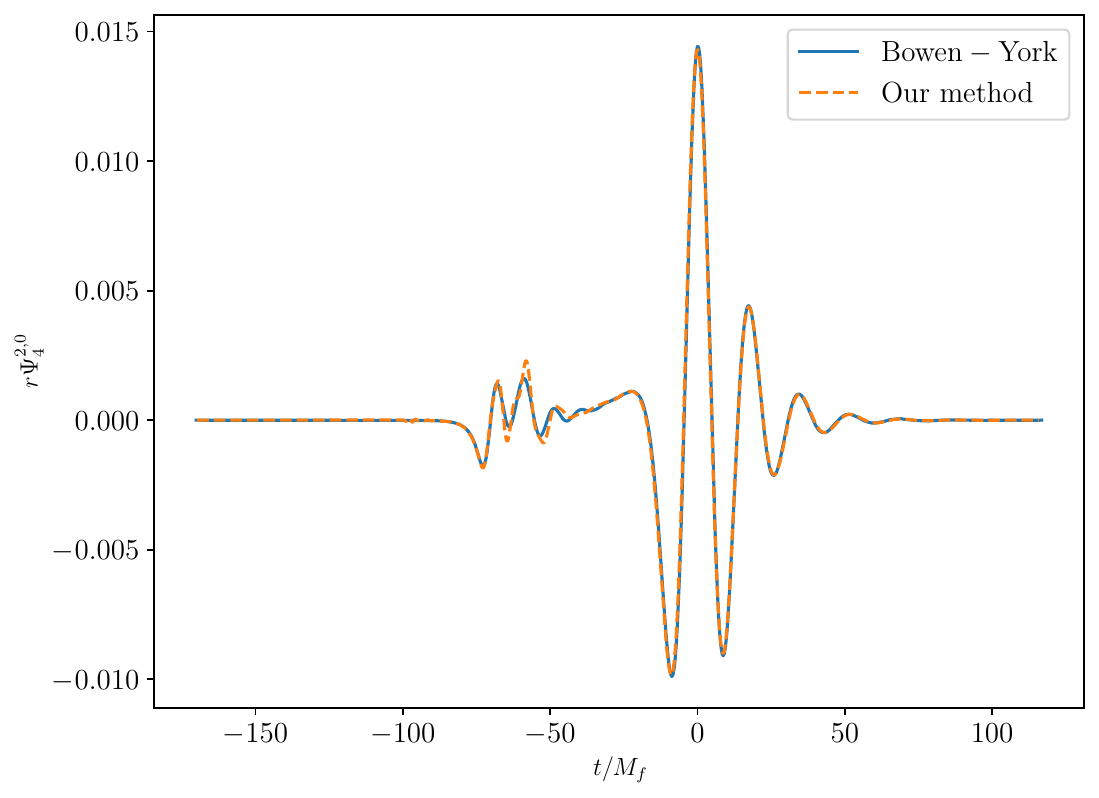}
    \includegraphics[height=5.5cm]{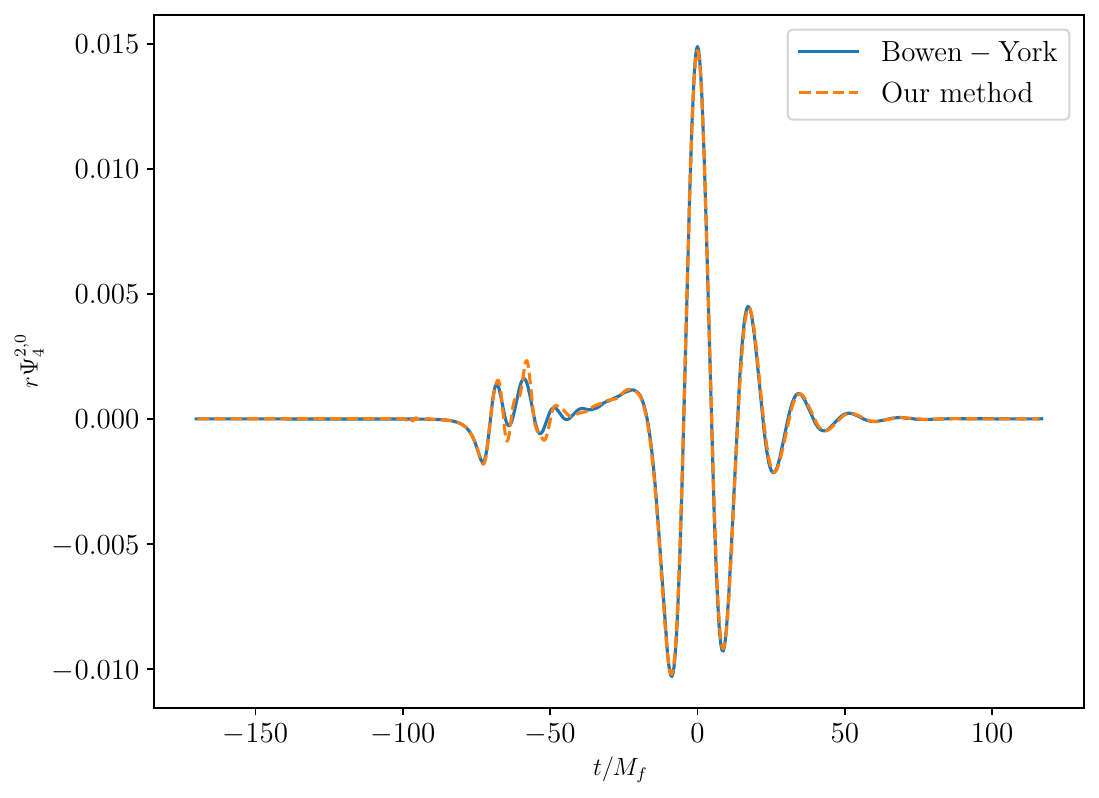} 
    \caption{The gravitational radiation $r\Psi_{4}^{20}$ from the head-on collision of equal-mass black holes produced by our initial data with the inner boundary condition for $\Omega$ prescribed as $\Omega|_{r=r_H}=\pm 82(3\cos^2{\theta}-1)$.
    Black holes have mass $M_1=M_2=0.5493$, and are initially placed at $z = \pm 6$.
    This is compared with Bowen-York black holes of the same masses and spins. Left: Both black holes have spins aligned along the $+z$ axis. Right: The black holes have opposite spin directions, aligned along the $+z$ and $–z$ axes, respectively.  }\label{fig_headon}
\end{figure*}

Next we set $\Omega|_{r=r_H}=\pm 82(3\cos^2{\theta}-1)$, which is, up to a constant, proportional to the spherical harmonic component $C_{20}$ (see the left panel in Fig.~\ref{C20_C30}). The $\pm$ sign corresponds to opposite spin directions.
Certain oblateness in the angular velocity is injected in the rotating horizon. $\bar K_P^{ij}$ is set to zero, so each black hole has no linear momentum initially. For the head-on collisions, no significant deviation from the Bowen-York waveform is observed, as shown in Fig.~\ref{fig_headon}.  

\subsection{Black hole Inspiral}

\begin{figure}[ht!]
    \includegraphics[height=6.1cm]{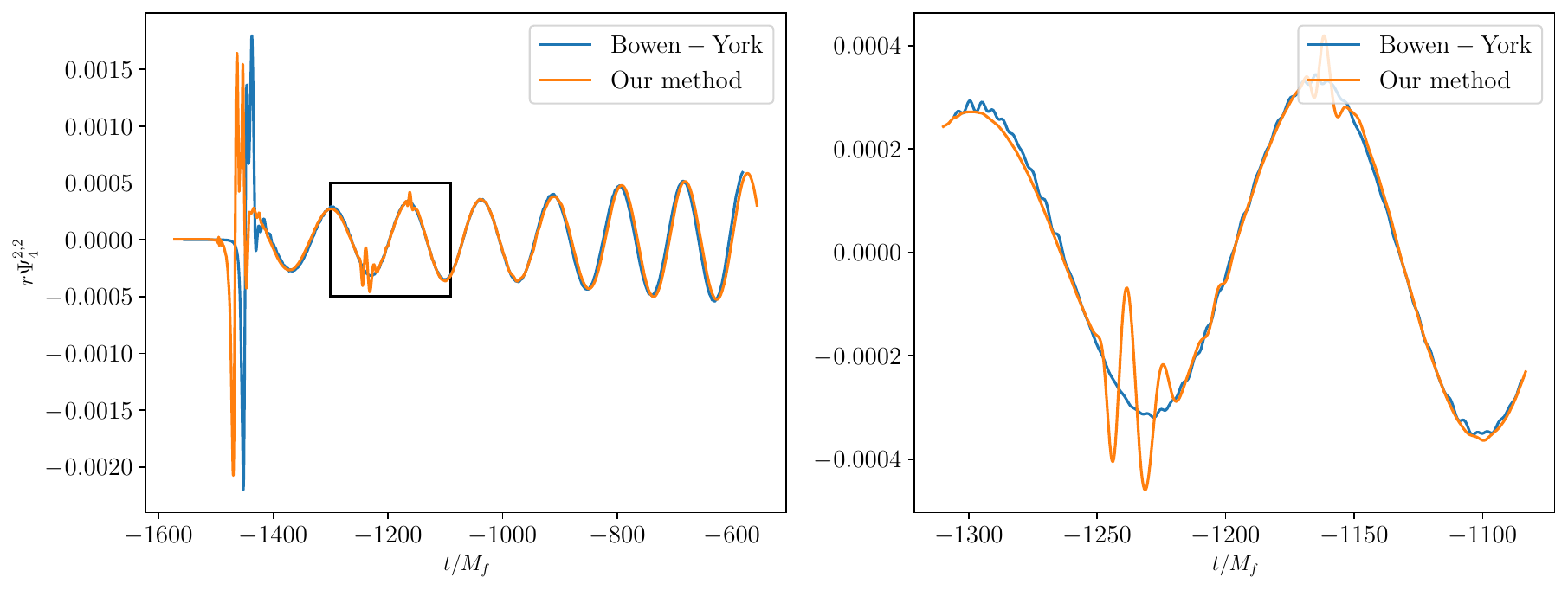}
	\caption{The gravitational radiation $r\Psi_{4}^{22}$ from the black hole inpiral of equal-mass, with aligned spin direction. The inner boundary conditions for $\Omega$ is $\Omega|_{r=r_H}=82(3\cos^2{\theta}-1)$. The black holes have masses $M_1=M_2=0.5511$, spins $S_1=S_2=(0.1312,0,0)$, and are initially placed at $x = \pm 6$. 
	\\In the left panel, the enclosed box highlights an oscillatory pattern during the early phase of the merger.
    The right panel provides a zoomed-in view of this pattern. }\label{Inspiral_aligned_20}
\end{figure}

\begin{figure}[ht!]
	\includegraphics[height=6.1cm]{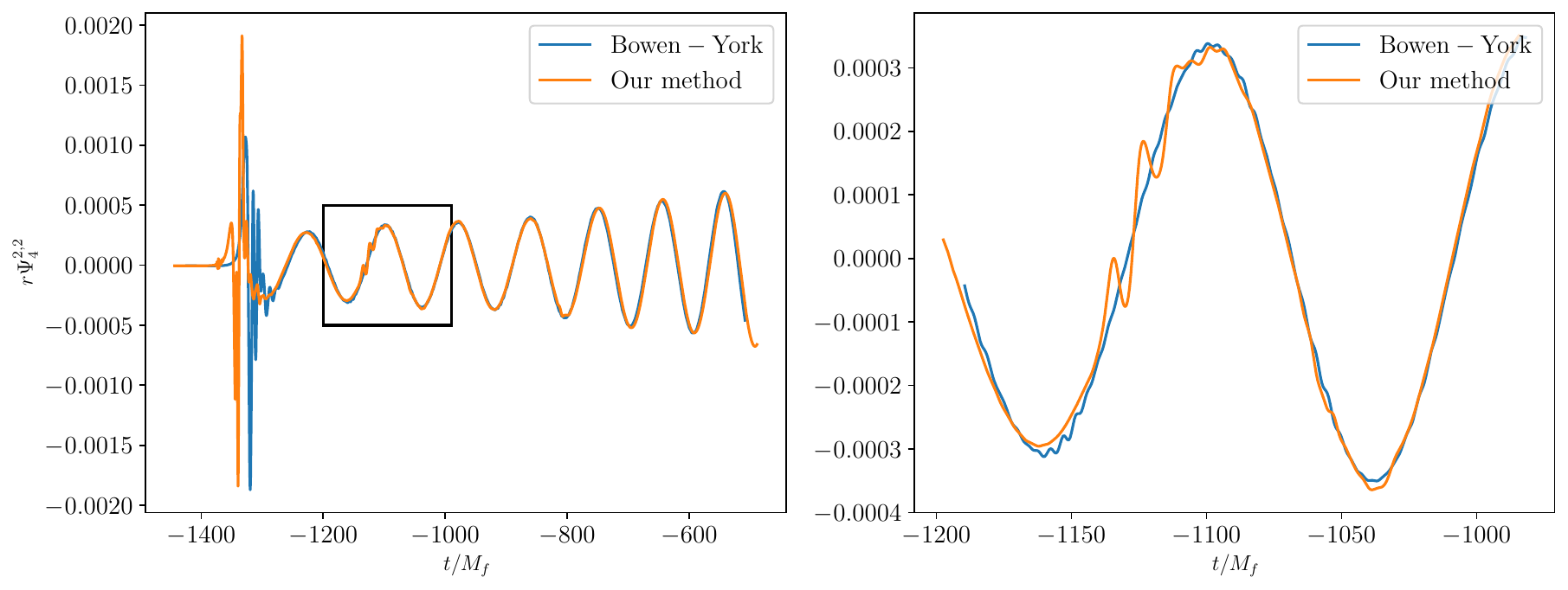}
	\caption{The gravitational radiation $r\Psi_{4}^{22}$ from the black hole inpiral of equal-mass, with anti-aligned spin direction.
	The inner boundary condition for $\Omega$ is $\Omega|_{r=r_H}=\pm 82(3\cos^2{\theta}-1)$. The black holes have masses $M_1=M_2=0.5511$, spins $S_1=(0.1312,0,0)$, $S_2=(-0.1312,0,0)$, and are initially placed at $x = \pm 6$. \\In the left panel, the enclosed box highlights an oscillatory pattern during the early phase of the merger.
    The right panel provides a zoomed-in view of this pattern.
	}\label{Inspiral_anti_20}
	\end {figure}

Next we activate $\bar K_P^{ij}$, and fine-tune the parameters such that the two black holes are initially in a nearly quasi-circular orbit. New features in the waveform begin to appear. As shown clearly in Fig.~\ref{Inspiral_aligned_20}, and \ref{Inspiral_anti_20}, the early-time waveform during the inspiral phase exhibits certain oscillatory pattern superposed on the overall inspiral waveform. 

The early-time waveform resembles a ringdown-like signal, reminiscent of the findings in \cite{bae2024ringdown}, which reported non-merging ringdown gravitational waves in simulations of hyperbolic black hole encounters. 
This connection supports the interpretation that the early ringdown-like features in our waveform are not gauge artifacts or residual junk radiation, but physically meaningful emissions triggered by the initial configuration of the black holes.

\begin{figure*}[ht!]
    \centering
    \includegraphics[height=5.5cm]{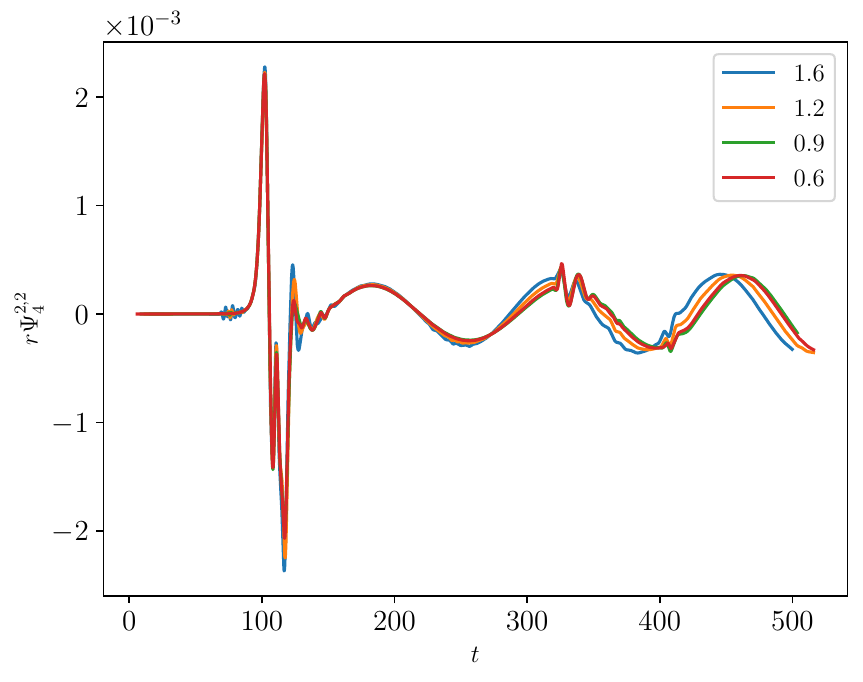}
    \includegraphics[height=5.5cm]{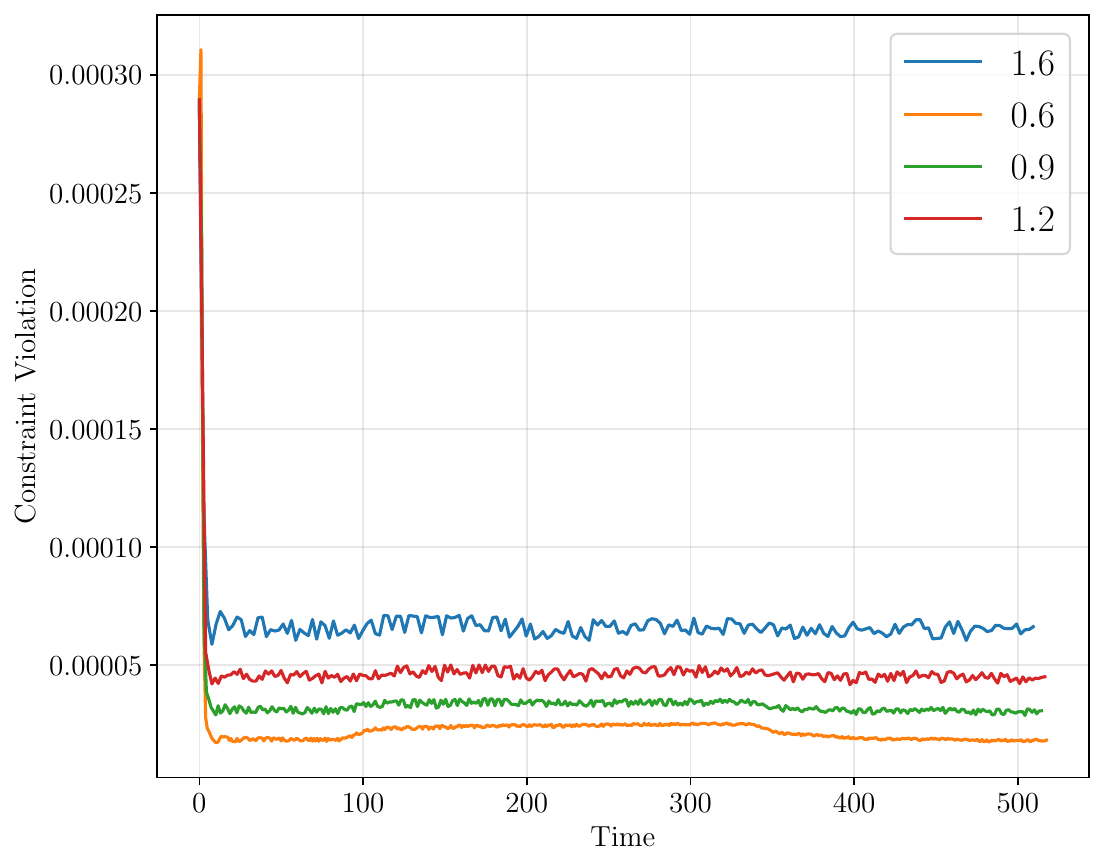} 
    \caption{Convergence test for the simulation shown in Fig.~\ref{Inspiral_aligned_20}.\\Left panel: Convergence of the waveform with increasing grid resolutions. Right panel: Systematic decrease of the Hamiltonian constraint violation with increasing grid resolution.}\label{fig_convergence}
\end{figure*}

To further confirm that these effects are physical rather than numerical artifacts, we perform a convergence test shown in Fig.~\ref{fig_convergence}. The left panel shows that the waveform converges as the resolution increases, with the phase exhibiting particularly strong agreement at higher resolutions. The right panel demonstrates that the Hamiltonian constraint violation decreases systematically with resolution, thereby confirming the consistency of the numerical simulation.

These results open the possibility that pre-merger QNM -like signatures, previously studied mostly in scattering or eccentric contexts, may also be present in quasi-circular binaries under certain initial conditions, though it is likely that the physical origins of the quasi-normal oscillatory pattern are different. 

\begin{figure}[ht!] 
\includegraphics[height=6.1cm]{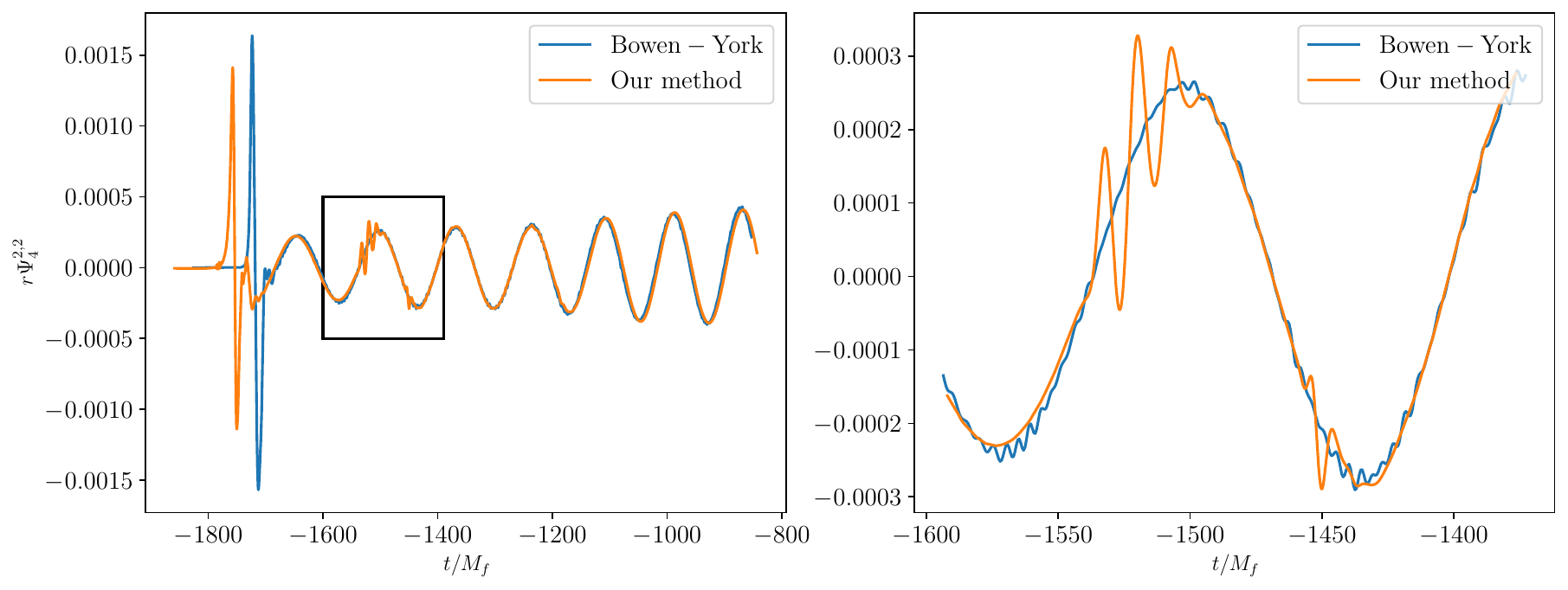}
\caption{The gravitational radiation $r\Psi_{4}^{22}$ from the black hole inpiral of equal-mass, with $\Omega|_{r=r_H}=80(5\cos^3{\theta} - 3\cos{\theta})$. The black holes have mass $M_1=M_2=0.5160$, spin $S_1=S_2=(10^{-8},0,0)$, and are initially placed at $x = \pm 6$. 
\\In the left panel, the enclosed box highlights an oscillatory pattern during the early phase of the merger.
The right panel provides a zoomed-in view of this pattern.
}\label{Inspiral_30}
	\end {figure}

\pagebreak
We further examine the case $\Omega|_{r=r_H}=80(5\cos^3{\theta} - 3\cos{\theta})$, in which the angular velocity is prescribed with a more complex profile corresponding to the spherical harmonic component $C_{30}$ (see the right panel in Fig.~\ref{C20_C30}). Again, the waveform in Fig.~\ref{Inspiral_30} displays an oscillatory pattern reminiscent of those of ringdown signals. 

\section{Concluding remarks} 

Based on a new geometric definition of the angular velocity of a rotating black hole, a new way to construct initial data for binary black holes is proposed. 
The new construction enables us to relate the geometry and dynamics at the horizon to the waveform at null infinity. The Hamiltonian and momentum constraints are formulated as a system of coupled elliptic equations for the conformal factor $\psi$ and the angular velocity $\Omega$, subject to appropriate boundary conditions at the horizons and at infinity. 

Waveform simulations in the conformally flat case using the new data suggest that, during the inspiral phase of binary black hole coalescence, new features of the waveform appears. This holds the promise  
that, in the detection of gravitational waves in space when many orbital cycles in the inspiral phase of black hole coalescence is expected to be observed, we will see new waveforms distinct from those currently observed by LIGO detection.
The present framework also allows the inclusion of matter in the constraint equations. The resulting initial data may then be used to probe the structure of neutron stars through their associated waveforms near null infinity. 
These problems remain to be studied in our future work.

\section*{Acknowledgements}

This work is dedicated to  Professor Ke Wu on the occasion of his 80th birthday, for his outstanding contributions to the development of mathematical physics in China. The
computations were partly done on the high performance computers of State Key Laboratory of Scientific and Engineering Computing, Chinese Academy of Sciences.
The work was supported by the National Key Research and Development Program of China (Grant No. 2021YFC2202501). The authors also acknowledge support from the National Natural Science Foundation of China (Grant No. W2431012).

\section*{References}

\bibliographystyle{unsrt}

\bibliography{refs.bib}

@article{feng2025geometric,
  title={Geometric inequality for axisymmetric black holes with angular momentum},
  author={Feng, Xuefeng and Yan, Ruodi and Gao, Sijie and Lau, Yun-Kau and Yau, Shing-Tung},
  journal={Classical and Quantum Gravity},
  volume={42},
  number={6},
  pages={065022},
  year={2025},
  publisher={IOP Publishing}
}

@book{kuhnau2002handbook,
  title={Handbook of Complex Analysis},
  author={Kuhnau, Reiner},
  year={2002},
  publisher={Elsevier}
}

@article{lovelace2008binary,
  title={Binary-black-hole initial data with nearly extremal spins},
  author={Lovelace, Geoffrey and Owen, Robert and Pfeiffer, Harald P and Chu, Tony},
  journal={Physical Review D},
  volume={78},
  number={8},
  pages={084017},
  year={2008},
  publisher={APS}
}

@article{bowen1980time,
  title={Time-asymmetric initial data for black holes and black-hole collisions},
  author={Bowen, Jeffrey M and York Jr, James W},
  journal={Physical Review D},
  volume={21},
  number={8},
  pages={2047},
  year={1980},
  publisher={APS}
}

@article{brown1993quasilocal,
  title={Quasilocal energy and conserved charges derived from the gravitational action},
  author={Brown, J David and York Jr, James W},
  journal={Physical Review D},
  volume={47},
  number={4},
  pages={1407},
  year={1993},
  publisher={APS}
}

@article{ashtekar2001mechanics,
  title={Mechanics of rotating isolated horizons},
  author={Ashtekar, Abhay and Beetle, Christopher and Lewandowski, Jerzy},
  journal={Physical Review D},
  volume={64},
  number={4},
  pages={044016},
  year={2001},
  publisher={APS}
}

@article{ashtekar2003dynamical,
  title={Dynamical horizons and their properties},
  author={Ashtekar, Abhay and Krishnan, Badri},
  journal={Physical Review D},
  volume={68},
  number={10},
  pages={104030},
  year={2003},
  publisher={APS}
}

@article{dain2002new,
  title={New conformally flat initial data for spinning black holes},
  author={Dain, Sergio and Lousto, Carlos O and Takahashi, Ryoji},
  journal={Physical Review D},
  volume={65},
  number={10},
  pages={104038},
  year={2002},
  publisher={APS}
}

@article{christodoulou1970reversible,
  title={Reversible and irreversible transformations in black-hole physics},
  author={Christodoulou, Demetrios},
  journal={Physical Review Letters},
  volume={25},
  number={22},
  pages={1596},
  year={1970},
  publisher={APS}
}

@article{shibata1995evolution,
  title={Evolution of three-dimensional gravitational waves: Harmonic slicing case},
  author={Shibata, Masaru and Nakamura, Takashi},
  journal={Physical Review D},
  volume={52},
  number={10},
  pages={5428},
  year={1995},
  publisher={APS}
}

@article{baumgarte1998numerical,
  title={Numerical integration of Einstein’s field equations},
  author={Baumgarte, Thomas W and Shapiro, Stuart L},
  journal={Physical Review D},
  volume={59},
  number={2},
  pages={024007},
  year={1998},
  publisher={APS}
}

@article{gourgoulhon2001quasiequilibrium,
  title={Quasiequilibrium sequences of synchronized and irrotational binary neutron stars in general relativity: Method and tests},
  author={Gourgoulhon, Eric and Grandclement, Philippe and Taniguchi, Keisuke and Marck, Jean-Alain and Bonazzola, Silvano},
  journal={Physical Review D},
  volume={63},
  number={6},
  pages={064029},
  year={2001},
  publisher={APS}
}

@article{grandclement2002binary,
  title={Binary black holes in circular orbits. II. Numerical methods and first results},
  author={Grandcl{\'e}ment, Philippe and Gourgoulhon, Eric and Bonazzola, Silvano},
  journal={Physical Review D},
  volume={65},
  number={4},
  pages={044021},
  year={2002},
  publisher={APS}
}

@article{bae2024ringdown,
  title={Ringdown gravitational waves from close scattering of two black holes},
  author={Bae, Yeong-Bok and Hyun, Young-Hwan and Kang, Gungwon},
  journal={Physical Review Letters},
  volume={132},
  number={26},
  pages={261401},
  year={2024},
  publisher={APS}
}

@Article{Ansorg:2004ds,
  requested-for ={EinsteinInitialData/TwoPunctures},                            
  author =       {Ansorg, Marcus and Br{\"u}gmann, Bernd and Tichy, Wolfgang}, 
  title =        {A single-domain spectral method for black hole puncture data}, 
  journal =      {Phys. Rev. D},                                                
  volume =       70,                                                            
  year =         2004,                                                          
  pages =        064011,                                                        
  eprint =       {arXiv:gr-qc/0404056},                                         
  doi =          {10.1103/PhysRevD.70.064011},                                  
  adsurl =       {https://www.slac.stanford.edu/spires/find/hep/www?rawcmd=find+doi+10.1103/PhysRevD.70.064011},                                               
  SLACcitation = "%%CITATION = GR-QC/0404056;%%"
}

@article{Schnetter:2003rb,                                                   
  requested-for ="Carpet/Carpet",    
  author =       "Schnetter, Erik and Hawley, Scott H. and Hawke, Ian",    
  title =        "{Evolutions in 3-D numerical relativity using fixed    
                  mesh refinement}",    
  journal =      "Class. Quantum Grav.",    
  volume =       21,    
  pages =        "1465-1488",    
  doi =          "10.1088/0264-9381/21/6/014",    
  year =         2004,    
  eprint =       "arXiv:gr-qc/0310042",    
}

@Article{Thornburg:2003sf,                                                   
  requested-for ="EinsteinAnalysis/AHFinderDirect",                          
  author =       "Thornburg, Jonathan",                                      
  title =        "{A Fast Apparent-Horizon Finder for 3-Dimensional          
                  Cartesian Grids in Numerical Relativity}",                 
  journal =      "Class. Quantum Grav.",                                     
  volume =       21,                                                         
  year =         2004,                                                       
  pages =        "743-766",                                                  
  eprint =       "arXiv:gr-qc/0306056",                                      
  adsurl =                                                                   
                  {https://www.slac.stanford.edu/spires/find/hep/www?rawcmd=find+doi+10.1088/0264-9381/21/2/026},                                         
  doi =          "10.1088/0264-9381/21/2/026",      
  SLACcitation = "%%CITATION = GR-QC/0306056;%%"
}

@Misc{EinsteinToolkit:2025_05,
  author       = {Maxwell Rizzo and Roland Haas and Steven R. Brandt and Zachariah Etienne and  Deborah Ferguson and Lucas Timotheo Sanches and Bing-Jyun Tsao and      Leonardo Werneck and David Boyer and Gabriele Bozzola and others},
  title        = {{The Einstein Toolkit. The "Martin D. Kruskal" release, ET\_2025\_05}},
  howpublished         = {\doi{10.5281/zenodo.15520463}},
}

@Article{Brown:2008sb,                                                                                           
     suggested-for ="Einsteinanalysis/Noexcision McLachlan/ML_BSSN",                                               
     author =       "Brown, J. David and Diener, Peter and Sarbach,                                              
                     Olivier and Schnetter, Erik and Tiglio, Manuel",                                            
     title =        "{Turduckening black holes: an analytical and                                                
                     computational study}",                                                                      
     journal =      "Phys. Rev. D",                                                                              
     volume =       "79",                                                                                        
     year =         2009,                                                                                        
     pages =        044023,                                                                                      
     eprint =       "arXiv:0809.3533 [gr-qc]",                                                                   
     adsurl =          {https://www.slac.stanford.edu/spires/find/hep/www?rawcmd=find+doi+10.1103/PhysRevD.79.044023},
     doi =          "10.1103/PhysRevD.79.044023",                                                                
     SLACcitation = "%%CITATION = 0809.3533;%%"                                                                  
}

@Misc{McLachlan:web,
  suggested-for ="McLachlan/ML_BSSN",
  key =          {McLachlan},
  title =        {{McLachlan}, a Public {BSSN} Code},
  howpublished = {\url{https://www.cct.lsu.edu/~eschnett/McLachlan/}},
}

@article{loffler2012einstein,
  title={The Einstein Toolkit: a community computational infrastructure for relativistic astrophysics},
  author={L{\"o}ffler, Frank and Faber, Joshua and Bentivegna, Eloisa and Bode, Tanja and Diener, Peter and Haas, Roland and Hinder, Ian and Mundim, Bruno C and Ott, Christian D and Schnetter, Erik and others},
  journal={Classical and Quantum Gravity},
  volume={29},
  number={11},
  pages={115001},
  year={2012},
  publisher={IOP Publishing}
}

@article{Dreyer:2002mx,
  requested-for ="LSUThorns/QuasiLocalMeasures",
  author =       "Dreyer, Olaf and Krishnan, Badri and Shoemaker,
                  Deirdre and Schnetter, Erik",
  title =        "{Introduction to isolated horizons in numerical
                  relativity}",
  journal =      "Phys. Rev. D",
  volume =       67,
  pages =        024018,
  doi =          "10.1103/PhysRevD.67.024018",
  year =         2003,
  eprint =       "arXiv:gr-qc/0206008",
}

@Misc{Lorene:web,
  title = {{LORENE}: Langage Objet pour la RElativité NumériquE},
  howpublished = {\url{https://lorene.obspm.fr/}},
}

\end{document}